\documentstyle[11pt]{article}
\title{\nopagebreak
\begin{flushright}
\tenrm UCTP109.00
\end{flushright}\vskip .7in
\large \bf  Twisted Kac-Moody Algebras \\  And \\
The Entropy Of ${\bf AdS}_3$ Black Hole}
\author{Sharmanthie Fernando$^{a}$ \thanks{e-mail address:
fernando@nku.edu} \,\,
and Freydoon Mansouri$^{b}$ \thanks{e-mail address:
mansouri@uc.edu} \\ 
\it \small 
$^{a}$Physics Department, Northern Kentucky University,
Highland Heights, KY 41099, USA, \\
\it \small
$^{b}$Physics
Department, University of
Cincinnati, Cincinnati, OH 45221, USA}
\date{}

\begin{document}
\maketitle

\begin{abstract}

We show that an $SL(2,R)_L \times SL(2,R)_R$ Chern-Simons theory
coupled to a source on a manifold with the topology of a disk
correctly describes the entropy of the ${\rm AdS}_3$ black hole.
The resulting boundary WZNW theory leads to two copies of a
twisted
Kac-Moody algebra, for which the respective Virasoro
algebras have the same central charge $c$ as the corresponding
untwisted theory. But the eigenvalues of the respective $L_0$ 
operators are shifted. We show that the asymptotic density of
states for this theory is, up to logarithmic corrections, the
same as that obtained by Strominger using the asymptotic symmetry
of Brown and Henneaux.

\end{abstract}
\pagebreak

\section{Introduction}
The entropy of the ${\rm AdS}_3$ black hole~\cite{btz,bhtz},
has been investigated from a
variety of points of view. Some of the more prominent approaches
to this problem have been compared and contrasted by
Carlip~\cite{carlip1}. In this
work we will address this problem in the framework of pure
Gravity in $2+1$ dimensions. Within this framework, a direct
method of obtaining the entropy of the BTZ black hole was given
by Strominger~\cite{andy}, in which use is made of the earlier
work of Brown and Henneaux~\cite{brown}. Using their results, he
demonstrated that 
the asymptotic symmetry of the BTZ black hole
is generated by two copies of the Virasoro algebra
with central charges
\begin{equation}
c_L = c_R = \frac{3l}{2G},
\end{equation}
where $l$ is the radius of curvature of the ${\rm AdS}_3$ space,
and $G$ is Newton's constant. Then, assuming that the ground
state eigenvalue $\Delta_0$ of the Virasoro generator $L_0$
vanishes, he obtained
the Bekenstein-Hawking
expression for the entropy. As pointed out by
Strominger~\cite{andy}, in this derivation one must take for
granted the existence
of a quantum gravity theory with appropriate symmetries. In the
absence of such a quantum theory, there will be no practical way
of computing either $\Delta_0$ or the value of the
classical central charge given by Eq. (1) from first principles.

Other approaches to the entropy problem make use of
the Chern-Simons theory representation of gravity in $2+1$
dimensions~\cite{achucarro,witten1}.
One common feature among them is that
to account for the microscopic degrees of freedom of the
black hole, the free Chern-Simons theory is formulated on a
manifold
with boundary~\cite{carlip2,bal,banados1}. Although the
significance of the boundary in these works differ, they all lead
to WZNW theories~\cite{witten2}.
More recently, these scenarios have been
further refined, improved, and
extended~\cite{carlip2,ortiz,banados2,banados3}.
One important feature of a typical conformal field theory 
obtained in this way is that its central charge varies between
the rank and the dimension of the gauge group. The relevant gauge
groups for the ${\rm AdS}_3$ black hole are two copies of the
group $SL(2,R)$, so that in the corresponding Virasoro algebras
the central charges vary in the range $1 \leq c \leq 3$. On the
other hand, the values of the central charges given by Eq. (1)
are very large and seem to be unrelated to Kac-Moody algebras
arising from relevant gauge groups. Thus, it appears that in the
Chern-Simons approach one reaches an impasse in providing a
quantum mechanical basis for the classical results of Brown and
Henneaux.

In this work, we describe a way to resolve this apparent
contradiction by interpreting the classical asymptotic Virasoro
algebra of Brown and Henneaux~\cite{brown} as an ``effective''
symmetry characterized by an ``effective central charge''in the
sense defined by Carlip~\cite{carlip1}. Then, rather than naively
comparing central charges, we derive the consequences of this
effective theory, including its ``effective central charge'' from 
yet another approach which makes use of 
Chern-Simons theory but which is physically very different from
the ones mentioned above. To begin with, in contrast to previous
works, in our approach the Chern-Simons theory is coupled to a
source. Then, since the BTZ black hole is a solution of
source-free Einstein's equations~\cite{btz,bhtz}, it is clear
that the
manifold $M$ on which the Chern-Simons theory is defined cannot
be identified with space-time. Instead, as shown in previous 
work~\cite{fm1,fm2}, the classical black hole space-time can be
constructed from the information encoded in the manifold $M$. In
particular, this information supplied, mass, angular momentum,
and the all important discrete identification
group~\cite{btz,bhtz} which distinguishes the black hole from
anti-de Sitter space.  One
important advantage of this point of view is that the manifold
$M$ is specified by its topology (no metric). As a result, for a
manifold with the topology of, say, a disk, the "size" of $M$ and
the location of the boundary relative to the source does not
enter
into the formalism, and a conformal field theory constructed on
its boundary is independent of where that boundary is. In other
words, it is unnecessary to specify whether the boundary refers
to a horizon or to asymptotic infinity.

Just as in obtaining the classical features of the black hole
space-time~\cite{fm1,fm2}, the coupling to a source turns out to
be
essential in arriving at a microscopic description of the black
hole entropy. In particular, it results in a conformal field
theory on the boundary with two copies of a {\it twisted} affine
Kac-Moody
algebra. In the corresponding Virasoro algebra, the value
of the central charge remains the same as the theory without a
source, but
the eigenvalues of the operator $L_0$ are shifted and are
non-vanishing. Taking these 
features as well as the subtleties that
arise from the non-compactness of $SL(2,R)$ into account, we find
that the asymptotic density of states for this microscopic theory
agree with that given by
Strominger~\cite{andy} if we identify the Brown-Henneaux values
for the central charge~\cite{brown} with the effective central
charge $c_{eff}$ of our theory.

\section{Chern-Simons Action and Boundary Effects}
For  a simple or a semi-simple Lie group, the Chern Simons action
has
the form
\begin{equation}
I_{cs} = \frac{k}{4\pi}Tr \int_M A \wedge \left( dA + \frac{2}{3}
A \wedge A\right) \end{equation}
where Tr stands for trace and
\begin{equation}
A = A_{\mu} dx^{\mu}
\end{equation} 
We require the 2+1 dimensional manifold M to have the topology 
$R\times\Sigma$, with $\Sigma$ a two-
manifold and $R$ representing the time-like coordinate $x^0$.
Moreover, we take the
topology of $\Sigma$ to be trivial in the absence of sources,
with the possible exception of a boundary.
Then, subject to the
constraints
\begin{equation}
F^b[A] =\frac {1}{2} \epsilon^{ij}
(\partial_i
A_j^{b} - \partial_j A_i^{b} + \epsilon^b_{\;cd}
A_i^{c}
A_j^{d}) = 0 \end{equation}
the Chern-Simons action for a simple group $G$ will take the form
\begin{equation}
I_{cs} = \frac{k}{2 \pi} \int_R dx^{0}  \int_{\Sigma}
d^2x\left(- 
\epsilon^{ij}\eta_{ab} A^{a}_i \partial_0 A^{b}_j + A^{a}_ 0
F_a \right)
\end{equation}
where $i,j = 1,2$.

We want to explore the properties of the Chern-Simons theory
coupled to a source for the group $SL(2,R)_L \times SL(2,R)_R$ 
on a
manifold with boundary. Since the gauge group is semi-simple, the
theory breaks up into two parts, one for each $SL(2,R)$, where by
$SL(2,R)$ we mean its infinite cover. So, to
simplify the presentation, we will study a single $SL(2,R)$.
Much of what we discuss in this and the next section hold for 
any simple Lie group, $G$. Also, to establish our
notation, we consider first
the theory in the absence of the source.

The main features of a Chern-Simons theory on a manifold with
boundary has been known for sometime~\cite{witten2,elitzur}.
Here, with $M= R \times \Sigma$, we identify the two dimensional
manifold $\Sigma$ with a disk $D$. Then, the boundary of $M$ will
have the topology $R \times S^1$. We parametrize $R$ with $\tau$
and
$S^1$ with $\phi$. 
In this parametrization, the Chern-Simons action on a manifold
with boundary can be written as
\begin{equation}
S_{cs}= \frac{k}{4 \pi} \int_{M} Tr ( AdA + \frac{2}{3} A^3) +
\frac{k}{4 \pi} \int_{\partial M} A_{\phi} A_{\tau}.
\end{equation}
The surface term vanishes in the gauge in which $A_{\tau}=0$ on
the boundary.
In this action, let $A = \tilde{A} + A_{\tau}$ and $d= d\tau
\frac{\partial}{\partial
\tau} + \tilde{d}$.
Then, the resulting constraint equations for the field strength
take the form
\begin{equation}
\tilde{F}= 0.
\end{equation}
They can be solved exactly by the ansatz~\cite{witten2,elitzur}
\begin{equation}
\tilde{A} = - \tilde{d}U U^{-1},
\end{equation}
where $U = U(\phi, \tau)$ is an element of the gauge group $G$.
Using this solution, the Chern-Simons action given by Eq. (5) can
be rewritten as
\begin{equation}
S_{WZNW}= \frac{k}{12 \pi} \int_M Tr ( U^{-1}dU)^3 + 
\frac{k}{4 \pi} \int_{\partial M} Tr(U^{-1}\partial_{\phi}U)
(U^{-1}\partial_{\tau}U)d\phi d\tau.
\end{equation}
We thus arrive at a WZNW action and can
take over many result already available in the literature for
this model. As in any WZNW theory, the change in the integrand of 
this action under an infinitesimal variation $\delta U$ of $U$ is
a derivative. We interpret this to mean that $U = U(\phi, \tau)$,
i.e., it is independent of the third (radial) coordinate of the
bulk. In other words, the information encoded in the disk depends
only on its topology and is invariant under any scaling of the
size of the disk.

The above 
Lagrangian is invariant under the following transformations
of the $U$ field~\cite{elitzur}:
\begin{equation}
U(\phi,\tau) \rightarrow \overline{\Omega}(\phi) U \Omega(\tau),
\end{equation}
where $\overline{\Omega}(\phi)$ and $\Omega (\tau)$ are any two
elements of $G$.
To obtain the conserved currents, let $ U \rightarrow U +
\delta U$. The corresponding variation of 
the action leads to
$ S_{WZNW} \rightarrow S_{WZNW} + \delta S_{WZNW} $, where
\begin{equation}
\delta S_{WZNW} = \frac{k}{2 \pi} \int_{\partial M} 
(\partial_{\tau}( U^{-1}
\partial_{\phi}U) \delta U. 
\end{equation}
This implies an infinite number of conserved currents:
\begin{equation}
J_{\phi} = -k U^{-1} \partial_{\phi}U = J^a_{\phi} T_a.
\end{equation}
Here, $T_a$ are the generators of the algebra $g$ of the group
$G$,
and
$J_{\phi}$ is a function of $\phi$ only because $\partial_{\tau}
J_{\phi}=0$.

If we expand $J_{\phi}$ in a Laurent series, we obtain
\begin{equation}
J_{\phi} = \Sigma J_n z^{-n - 1},
\end{equation}
where $z= exp(i \phi)$. As usual, $J_n$ satisfy  the Kac-Moody
algebra
\begin{equation}
[J^a_n,J^b_m] = f^{ab}_c J^c_{m+n} +  k n g^{ab}
\delta_{m+n,0}
\end{equation}
The corresponding energy momentum tensor for the action
$S_{WZNW}$
can be
computed using the Sugawara-Sommerfield
construction. For example, for the gauge group $SL(2,R)$,
\begin{equation}
T_{\phi \phi} = \frac{1}{(k - 2)} g_{ab}
{\large :} J_{\phi}^a (z) J_{\phi}^b (z) 
{\large :} = 
\frac{1}{k - 2} \Sigma {\large :} J^a_{n-m} J_m^a {\large :}
z^{-n-2} = \Sigma L_n z^{-n-2},
\end{equation}
where
\begin{equation}
L_n = \frac{1}{k - 2} \Sigma {\large :} J^a_{n-m} J_m^a {\large
:}.
\end{equation}
The $L_n$ operators satisfy the following Virasoro algebra:
\begin{equation}
[L_n,L_m] = (n-m) L_{n+m} + \frac{c}{12} n(n^2-1) \delta_{n+m,0},
\end{equation}
with $c$ the central charge. For $SL(2,R)$, it is given by $c =
\frac{3k}{k - 2}$. We note that
for large negative values of $k$, the value of $c$ approaches $3$
which is the dimension of the group. 
We also note that this boundary WZNW theory has
one, not the more usual two, Virasoro algebra. It will
be shown below that when the Chern-Simons theory is coupled to a
source on a manifold with the topology of a disk, the central
charge of the Virasoro algebra of the corresponding modified WZNW
theory
remains the
same as that in the source-free theory discussed above.

\section{The Coupling of a source}
Next, we couple a source to the Chern-Simons action on the
manifold $M$ with disk topology, which, as in the previous
section, has the boundary
$R \times S^1$. In general, we take the source to be
a unitary representation of the group $G$. To be more specific,
let us 
consider a source action given by~\cite{witten2,elitzur}
\begin{equation}
S_{source} =  \int d\tau Tr[ \lambda \omega(\tau)^{-1} (
\partial_{\tau} +
A_{\tau}) \omega(\tau)].
\end{equation}
Here $\lambda = \lambda^i H_i$, where $H_i$ are elements of
the Cartan subalgebra $H$ of $G$. We will take $\lambda^i$ to be
appropriate weights.
The quantity $\omega (\tau)$ is
an arbitrary element of $G$. The above action is invariant
under the transformation $\omega(\tau) \rightarrow \omega(\tau)
h(\tau)$,
where $h(\tau)$ commutes with
$\lambda$.
 
Now the total action on $M$ is,
\begin{equation}
S_{total}= \frac{k}{4 \pi} \int Tr ( AdA + \frac{2}{3} A^3) + 
\frac{k}{4 \pi} \int A_{\tau} A_{\phi}
+   \int d\tau Tr[ \lambda \omega(\tau)^{-1} ( \partial_{\tau} +
A_{\tau})
\omega(\tau)]
\end{equation}
The new constraint equation takes the form,
\begin{equation}
\frac{k}{2 \pi} \tilde{F}(x) + \omega(\tau) \lambda
\omega^{-1}(\tau)
\delta^2(x-x_p) = 0,
\end{equation}
where $x_p$ specifies the location of the source, heretofore
taken to be at $x_p = 0$.
The solution to the above equation is given by
\begin{equation}
\tilde{A} = - \tilde{d}\tilde{U} {\tilde{U}}^{-1},
\end{equation}
where~\cite{elitzur}
\begin{equation}
\tilde{U} = U exp (\frac{1}{k}\omega(\tau) \lambda
\omega^{-1}(\tau)\phi)
\end{equation}
The new effective action on the boundary $\partial M$ is then
\begin{equation}
S_{total} = S_{WZNW} + \frac{1}{2 \pi} \int_{\partial M} Tr (
\lambda U^{-1} \partial_{\tau} U).
\end{equation}
This Lagrangian is
also invariant
under the following transformation:
\begin{equation}
U(\phi,\tau) \rightarrow \overline{\Omega}(\phi) U \Omega(\tau)
\end{equation}
where $\Omega(\tau)$ commutes with $\lambda$. Varying the action
under the above symmetry transformation, we get
\begin{equation}
\delta S_{total} = \delta S_{WZNW} + \delta S_{source},
\end{equation}
where
\begin{equation}
\delta S_{source} = \frac{1}{2\pi} \int Tr \left( -U^{-1} \delta
U [ U^{-1}\partial_{\tau}U, \lambda] \right).
\end{equation}
Hence, the requirement that $\delta S_{total} = 0$ will give rise
to the conservation equation~\cite{afm}
\begin{equation}
\partial_{\tau} \left( - k{U}^{-1}
\partial_{\phi}{U} \right) + [ {U}^{-1} \partial_{\tau}{U},
\lambda] = 0.
\end{equation}
The first term in this expression has the same structure as the
current $J_{\phi}$ of the source free theory. Hence, requiring
that $U(\phi, \tau) = U(\phi + \tau)$, we can write
the new current $\hat{J}_{\phi}$ in terms of the
current in the absence of the source as
\begin{equation}
\hat{J}_{\phi} = e^{\frac{\lambda}{k} (\phi + \tau)} 
J_{\phi}
e^{-\frac{\lambda}{k} (\phi + \tau)}
\end{equation}
It is easy to check that
\begin{equation}
\partial_{\tau} \hat{J}_{\phi} = 0
\end{equation}

With the new currents at our disposal, the next step is to see
how this modification affects the properties of the corresponding
conformal field theory. In this respect, we note from Eq. (28)
that our new currents $\hat{J}_{\phi}$ are related to the
currents $J_{\phi}$ in the absence of the source by a conjugation
with respect to the elements of the Cartan subalgebra $H$ of the
group $G$. This kind of conjugation has been noted in the study
of Kac-Moody algebras~\cite{olive,suranyi,halpern}: The algebra
satisfied by the new currents fall in the category of {\it
twisted} affine Kac-Moody algebras.
So, to understand how the
coupling to a source modifies the structure of the source-free
conformal field theory, we follow the analysis of
reference~\cite{olive} and express the Lie algebra of the group
$G$ of rank $r$ in the Cartan-Weyl basis.
Let $H^i$ be the elements of the Cartan subalgebra and denote the
remaining generators by $E^{\alpha}$. Then, with label $a = (i,
\alpha )$,
\begin{equation}
[H^i, H^j] = 0 \, ; \hspace{1cm}
[H^i, E^{\alpha}] = \alpha^i E^{\alpha}
\end{equation}
\begin{eqnarray}
[E^{\alpha}, E^{\beta}] = \left\{ \begin{array}{cc} 
\epsilon(\alpha, \beta) E^{\alpha + \beta} \hspace{0.3cm}
\mbox{if $ \alpha + \beta $ is a root } \\
2 \alpha^{-2}( \alpha .H) \hspace{0.3cm}
\mbox{if $  \alpha = - \beta$}\\
0 \hspace{0.5cm} \mbox{otherwise}
\end{array} \right\}
\end{eqnarray}
In this expression, $1 \leq i,j \leq r$, and $\alpha, \beta$ are 
roots. Now we can rewrite the affine Kac-Moody algebra $g$ of the
source free theory of the last section in this basis
as follows:
\begin{equation}
[H^i_m, H^j_n] = k m \delta^{ij} \delta_{m,-n}\, ; \hspace{0.5cm}
[H^i_m, E^{\alpha}_n] = \alpha^i E^{\alpha}_{m+n}
\end{equation}
\begin{eqnarray}
[E^{\alpha}_m, E^{\beta}_n] = \left\{ \begin{array}{cc} 
\epsilon(\alpha, \beta) E^{\alpha + \beta}_{m+n} \hspace{0.4cm}
\mbox{if $ \alpha + \beta $ is a root} \\
2 \alpha^{-2}(  \alpha .H_{m+n} + km \delta_{m, -n})
\hspace{0.3cm}
\mbox{if $  \alpha = - \beta $} \\
0 \hspace{0.6cm} \mbox{ otherwise}
\end{array}  \right\}
\end{eqnarray}
We also note from the last section that in the absence of
the source the element $L_0$ of the Virasoro algebra of the
source-free theory and the
currents $J^a_n$ have the following commutation relations:
\begin{equation}
[L_0, J^a_n] = - n J^a_n.
\end{equation}

It can be seen from Eq. (28) that the
new currents can be viewed
as an inner automorphism of the algebra $g$ in the form 
$\zeta(J) = 
\gamma J \gamma^{-1}$. The effect of this on the component
currents
can be represented by
\begin{equation}
 e^{ i \chi. H}
\end{equation}
As a result of this inner automorphism on elements of the algebra
$g$, we obtain a
modified algebra $\hat{g}$ the elements of which in the
Cartan-Weyl basis are given by~\cite{olive}
\begin{equation}
\zeta(H^i) = H^i \,;
\hspace{1.0cm}
\zeta(E^{\alpha}) = e^{i \chi . \alpha} E^{\alpha}
\end{equation}
If the map $\zeta$ is endowed with the property that 
$\zeta^N = 1$, then we must have
 $N \chi.\alpha = 2n \pi$, where $n$ is an integer
for all roots $\alpha$\, $ \epsilon$\, $g$ \, :
\begin{equation}
e^{i \chi . \alpha} = e^{\frac{2 \pi in}{N}}.
\end{equation}
where $n$ is a positive integer $\leq N - 1$. 
As far as the currents obtained from the Chern-Simons theory
coupled to a source are concerned, it appears that all possible
values of $N$ are
allowed. However, for algebras of low rank such as, the value of
$N$ can be unique. This is because
the automorphism $\zeta$ divides a suitable
combination of the generators of $\hat{g}$ into eigenspaces
$\hat{g}_{(m)}$. For this arrangement to work for $SU(2)$ or
$SL(2,R)$, the only non-trivial possibility is $N = 2$.

Thus, the  basis of $\hat{g}$ consists of the elements
$H^i_m$ and $E^{\alpha}_n$
where $m$ $ \epsilon$ $ Z$ and $n$ $ \epsilon$ $ (Z +
\frac{\chi . \alpha }{2 \pi})$.
These operators satisfy a Kac-Moody algebra which has formally
the same structure
as that of $g$ but with rearranged (fractional) values of the
suffices.
Hence the algebra $\hat{g}$
can be viewed as the ``twisted'' version of the  algebra
$g$.

Since the automorphism which relates the two algebras is of inner
variety~\cite{olive}, we must look for features, if any, that
distinguish
the algebra $\hat{g}$ from its untwisted version $g$.
These features depend on the extent to which we can
undo the twisting. To this end, we
introduce a new basis for $\hat{g}$
\begin{equation}
\hat{E}^{\alpha}_n = E^{\alpha}_{ n + \frac{\chi . \alpha} 
{2 \pi}}; \hspace{1.0cm} \hat{H}^i_{n} 
= H^i_{n} + \frac{k}{2 \pi} \chi^i  \delta_{n,0},
\end{equation}
The new operators, $\hat{E}^{\alpha}_{n}$ and $\hat{H}^i_{n}$
satisfy the same commutation relations as the elements of the 
untwisted affine Kac-Moody algebra $g$.
The corresponding conformal field theories are not identical,
however. This can be seen most easily if we express the Virasoro
generators $\hat{L}_n$ of the twisted theory in terms of
untwisted generators:
\begin{equation}
\hat{L}_m = L_m - \frac{1}{2 \pi} \chi^i H^i_{n} + 
\frac{k}{4 \pi^2} \chi^{i} \chi^{i} \delta_{n,0}.
\end{equation}
In particular, we get for
$\hat{L}_0$,
\begin{equation}
\hat{L}_0 = L_0 - \frac{1}{2 \pi} \chi^{i} H^i_{0} + \frac{k}
{4 \pi^2} \chi^{i} \chi^{i}. 
\end{equation}
Thus the eigenvalues $\hat{\Delta}$ of the operator
$\hat{L}_0$ are shifted relative to the eigenvalues $\Delta$ of
$L_0$. But, as can be verified directly, the value of the central
charge $c$ remains unchanged~\cite{olive,suranyi,halpern}. More
specifically, we have
\begin{equation}
\hat{L}_0 |\hat{\Delta}, \mu> = \hat{\Delta} |\hat{\Delta}, \mu>,
\end{equation}
where $\mu$ is a weight and
\begin{equation}
\hat{\Delta} = \Delta - \frac{1}{2 \pi}\, \chi^{i} \mu^i + 
\frac{k}{4 \pi^2} \chi^{i} \chi^{i}.
\end{equation}
So, for the highest (lowest) weight states, we get
\begin{equation}
\hat{\Delta}_{0} = \Delta_0 - \frac{1}{2 \pi}
\chi^{i} \mu_0^i + \frac{1}{4 \pi^2}k
\chi^2.
\end{equation}

With minor exceptions, most of the derivation of our twisted Kac-
Moody algebras from the Chern-Simons theory applies to any gauge
group. But
the relation between the irreducible representations of
an affine Kac-Moody algebra and its Lie subalgebra imposes
restrictions on the value of the central term $k$. For example,
for $SU(2)$, the value of $k$ is restricted
to the non-negative values~\cite{olive}. But for discrete unitary
representations of $SL(2,R)$ with a lowest weight, the quantity
$k$ is restricted
to~\cite{sl2rkm}  
\begin{equation}
k < -1.
\end{equation}
It follows that in this case large negative values of $k$ are
allowed. 
We will take advantage of this feature in the application of this
formalism to
entropy of the 
${\rm AdS}_3$ black hole in the next section.

\section{The Entropy of the ${\bf AdS}_3$ Black Hole}
As pointed out in the introduction, in the derivation of the 
entropy of the ${\rm AdS}_3$ black hole by
Strominger~\cite{andy},
use was made of the expression for the central charges of the two
asymptotic Virasoro algebras obtained
by Brown and Henneaux~\cite{brown} using classical (non-quantum)
arguments. 
They are given by
\begin{equation}
c_L = c_R = \frac{ 3l}{2G}.
\end{equation}
where $l$ is the radius of curvature of the ${\rm AdS}_3$ space,
and $G$ is Newton's constant. The presence of such a symmetry
indicates that there is a conformal field theory at the
asymptotic boundary. It was shown by Strominger that the BTZ
solution satisfies the Brown-Henneaux boundary conditions so that
it possessed an asymptotic symmetry of this type. So, he
identified
the degrees of freedom of the black hole in the bulk with those
of the conformal field theory at the infinite boundary. Then,
using Cardy's formula~\cite{cardy} for the asymptotic density of
states, he showed that for $l \gg G$
the entropy of this conformal field theory
is given by
\begin{equation}
S\,=\, \frac{2\pi\,r_+}{4G},
\end{equation}
in agreement with Bekenstein-Hawking formula. Here, the quantity
$r_{+}$ is the outer horizon radius of the black hole. An
important assumption in this derivation was that the ground state
eigenvalue $\Delta_0$ of the operator $L_0$ vanishes.

The formula by Cardy~\cite{cardy} for the asymptotic density of
states, leading to the above expression for entropy is given by
\begin{equation}
\rho(\Delta) \approx exp\{2 \pi \sqrt{ \frac{c
\Delta}{6}}\},
\end{equation}
where $\rho (\Delta)$ is the number of states for
which the eigenvalue of $L_0$ is $\Delta$. The result holds when
$\Delta$ is large and
the lowest eigenvalue $\Delta_0$ vanishes. From the analysis of
the previous section, it is clear that in the conformal field
theory arising from a Chern-Simons theory coupled to a source
the eigenvalue $\hat \Delta_0$ does not vanish, so that the above
Cardy formula must be appropriately modified. In such a case, 
the asymptotic density of
states for large $\Delta$ is given by~\cite{carlip1}:
\begin{equation}
\rho(\Delta) \approx 
exp \{ 2 \pi \sqrt{\frac{(c - 24
\Delta_0)\Delta}{6}}
\}
\rho(\Delta_0) =
exp \{ 2 \pi \sqrt{\frac{c_{eff}\Delta}{6} } \}
\rho(\Delta_0).
\end{equation}
Thus, it is the latter formula which must be used in the
application of our formalism to the black hole entropy. It is
important to note that the expression for the asymptotic density
of states given by Eq. (48) rests on the existence of a
consistent conformal field theory with a well defined partition
function. For Kac-Moody algebras based on compact Lie groups this
can be established rigorously. But for Kac-Moody algebras based
on non-compact groups such as $SL(2,R)$, no general proof exists.
So, all the conformal field theories based on $SL(2,R)$, which
have been made use of in connection with the ${\rm AdS}_3$ black
hole, including the present work, share this common weakness.
 
We want
to show that the results obtained from such a microscopic
analysis are in agreement with those given by
Strominger~\cite{andy}. In so doing, we will rely heavily on our
previous results which dealt with the understanding of the 
macroscopic features of the BTZ black hole~\cite{fm1,fm2}.
We recall from these references that the unitary representations
of $SL(2,R)$ which are relevant to the description of the
macroscopic features of the black hole are the infinite
dimensional discrete series which are bounded from below. These
irreducible representations are characterized by a label $F$
which
can be identified with the lowest eigenvalue of the $SL(2,R)$
generator which is being diagonalized. In the literature of the
$SL(2,R)$ Kac-Moody algebra~\cite{sl2rkm}, this (lowest weight)
label, which is convenient for the description of this series,
is often referred to as $-j $. Thus, the Casimir eigenvalues
in the two notations are related according to
\begin{equation}
j(j +1) = F(F - 1),
\end{equation}
where $F \geq 1$.
On the other hand, in the description of the black holes in terms
of a Chern Simons theory with gauge group $SL(2,R)_L \times
SL(2,R)_R$, the mass and the angular momentum of the black hole
are related to the Casimir invariants of $j^2_{\pm}$ of this
gauge group as follows~\cite{fm1,fm2}:
\begin{equation}
j^2_{\pm} = \frac{1}{4} (lM \pm J).
\end{equation}
So, from Eqs. (49) and (50) we get for positive mass black holes
\begin{equation}
F_{\pm} = \frac{1}{2} \left[ 1 + \sqrt{ 1 + (lM \pm J)} \right].
\end{equation}
In particular~\cite{fm1,fm2}, 
for the lowest massive state, we must have $F^{(0)}_{\pm} \approx
1$. This means that
$(lM_0 + J_0) \ll 1$. We will assume that this dimensionless
quantity is proportional to the
ratio of the two scales of the theory:
\begin{equation}
(lM_0 + J_0) \approx \frac{2 \beta^2 G}{l}.
\end{equation}
Here $\beta$ is a real positive number, and the factor 2 is put
there for later convenience. The value of $\beta$ will be fixed
by requiring consistency between the classical and the quantum
black hole descriptions.

Next, consider the determination of the Chern-Simons couplings
$k_{\pm}$. These were referred to as $a_{\pm}^{-1}$ in
references~\cite{fm1,fm2}. Since the gauge group $SL(2,R)_L
\times
SL(2,R)_R$ is semi-simple, the couplings $k_{+}$ and $k_{-}$ are
independent. Also, since the two $SL(2,R)$ groups play a parallel
role
in our approach, we will focus on determining one of them, say,
$k_{+}$. The other one can be obtained in a similar way. To this
end, we recall from references~\cite{fm1,fm2} that in our
approach, the manifold $M$ on which the Chern-Simons theory is
defined is not space-time. This means that from the data encoded
in $M$ we must be able to obtain all the features of the
classical black hole space-time. One of these is the discrete
identification
group of the black hole~\cite{btz,bhtz}. The elements of this
group were obtained within
our framework by considering the holonomies around the source in
$M$. In this respect, we note that the Cartan sub-algebra of
$SL(2,R)$ is one
dimensional, and the corresponding weight is $F$ as discussed
above. Then, directly or using
non-abelian
Stokes theorem~\cite{kmr} and equation (20), the holonomies can
be
evaluated. To get the correct discrete identification group, this
implies that~\cite{fm1,fm2}
\begin{equation}
\frac{2F^{(0)}_{+}}{k_{+}} = \pm \frac{1}{l} (r_{+} + r_{-}) = 
\pm 2 \sqrt{ 2 \frac{G}{l} (lM_0 + J_0)},
\end{equation}
where $r_{+}$ and $r_{-}$ are, respectively, the outer and the
inner horizon radii of the black hole. A similar analysis can be
carried out for $k_{-}$. Using the values of $F^{(0)}$
and $(lM_0 + J_0)$ for the ground state given above, we find for
both couplings
\begin{equation}
k_{+} = k_{-} = \pm \frac{l}{2 \beta G}.
\end{equation}
The sign of $k_{\pm}$ is not fixed by holonomy considerations
alone. But from the discussion of the last section it is clear
that we must
choose the
negative sign for both since $SL(2,R)$ is non-compact.

Next, consider the ground state eigenvalue $\hat{\Delta}_0$
for one of the two $SL(2,R)$ Virasoro algebras. Since the Cartan
subalgebra is one dimensional, the sums in Eq. (43) consist of
one term each, and, from Eq. (37), the quantity $\chi$ is given
by
\begin{equation}
\chi \: = \: \frac{2 \pi n}{N \alpha}.
\end{equation}
The quantity $\mu_0$ in Eq. (43) is clearly the weight of the
ground state, i.e., the weight $F \approx 1$ described above. We
also note that the root $\alpha$ is the weight of the adjoint
representation of $SL(2,R)$, so that $\alpha = 1$. Moreover, as
discussed in the last section, for $SL(2,R)$ we have $\frac{n}{N}
= \frac{1}{2}$
Then, Eq. (44) specialized to the case at hand will take the form
\begin{equation}
\hat{\Delta}_0 = \Delta_0  - \frac{1}{2} + \frac{k}{4}.
\end{equation}
Also substituting for $k$ from Eq. (54), we get
\begin{equation}
\hat{\Delta}_0 = \Delta_0 - \frac{1}{2} 
- \frac{l}{8 \beta G}.
\end{equation}
In this expression, the quantity $\Delta_0$ is the ground state
eigenvalue of $L_0$, and its value is not known but is often
taken to be zero without{\it \'a priori} justification. For our
purposes, it is only necessary that it be small compared to the
last term.

Let us now compute the effective central charge $c_{eff}$,
defined via Eq. (48), for our theory. It is given by
\begin{equation}
c_{eff} = c - 24 \hat{\Delta}_0 = c -24 \Delta_0 
- 12 + \frac{3l}{\beta G}.
\end{equation}
In this expression, $\frac{l}{G} \gg 1$ whereas $ 1 \leq c \leq
3$. Assuming
that $\Delta_0$ is also relatively small, we get
\begin{equation}
c_{eff} \approx \frac{3l}{\beta G}.
\end{equation}

To determine the quantity $\beta$, we note that our
starting point, i.e., the Chern-Simons theory coupled to a source
leads, on the one hand, to the twisted Kac-Moody algebras and
conformal field theories
described above and, on the other hand~\cite{fm1,fm2}, to the
classical BTZ solution~\cite{btz,bhtz} for which the central
charge for the asymptotic Virasoro algebra was given by Brown and
Henneaux~\cite{brown}. It is then necessary that
the classical and quantum results which follow from the same
Chern-Simons theory be consistent with each other. So, it is
reasonable to require that the asymptotic density of states of
the above quantum theory, as computed from Cardy-
Carlip~\cite{carlip1}, agree, up to logarithmic terms, with the
asymptotic density of states obtained by Strominger~\cite{andy}
using the traditional Cardy formula and the classical Brown-
Henneaux value of central charge (45). The simplest way to
satisfy this requirement is to set our effective central charge
$c_{eff}$ equal to the Brown-Henneaux central charge. This fixes
$\beta = 2$. That this requirement makes sense
can also be seen by noting that for the Virasoro algebra obtained
by Brown and Henneaux, the underlying Kac-Moody algebra is not
known, so that there is no direct way of calculating its central
charge $c$ or its ground state eigenvalue from a fundamental Kac-
Moody algebra. Then, taking the classical theory to be an
``effective theory'', we see that we can compute its ``effective
central charge'' and its ``effective ground state eigenvalue''
from the above quantum theory up to a proportionality constant.
Fixing the value of $\beta$ in our approach using consistency is
to be compared with the fixing of the ground state eigenvalue
$\Delta_0$ by Strominger~\cite{andy}. In that work, the conformal
field theory alone does not limit the continuous infinity of the
possible values of $\Delta_0$, and the requirement that it vanish
has only {\it \'a posteriori} justification.

With $\beta = 2$, Eq. (54) implies that 
$k_{\pm} = - \frac{3l}{2G}$. Coincidentally, the magnitudes of
these quantities are the same as those obtained by another
approach~\cite{banados2} in which the manifold $M$ was taken to
be space-time and the {\it free} Chern-Simons action led to the
classical black hole solution in $M$. In that case, the signs of
$k_{+}$ and $k_{-}$ are opposite
each other, so that one of them would have to be positive. This
appears to be inconsistent with what we know about $SL(2,R)$ Kac-
Moody algebras.

The main result of this section is that for this conformal field
theory, the expression
for the asymptotic density of states given by Eq. (48) reduces to
\begin{equation}
\rho(\hat{\Delta}) \approx 
exp \{ 2 \pi \sqrt{\frac{c_{eff}\, \hat{\Delta}}{6} } \}
\rho(\hat{\Delta}_0) = exp \{ 2 \pi 
\sqrt{\frac{l\, \hat{\Delta}}{4G}} \}\, \rho( \hat{\Delta}_0).
\end{equation}
Modulo a logarithmic correction, this expression is identical
with that
used by Strominger~\cite{andy}. This resolves the longstanding
controversy in the traditional method of comparing the central
charge of a conformal field theory obtained from the Chern-Simons
approach with the (effective) classical results of Brown and
Henneaux. One of the crucial features of our work which led to
this resolution was the recognition that for $SL(2,R)$ Kac-Moody
algebras, large negative values of $k$ are allowed.

So far, we have dealt with the
density of states for one $SL(2,R)$, say, $SL(2,R)_L$. This will
contribute an amount $S_L$ to the black hole entropy.
Clearly, we can repeat this
computation for the density of states of $SL(2,R)_R$.
Then, to the extent that the logarithmic corrections can be
neglected, the black hole entropy $S = S_L + S_R$ is in agreement
with that given
by Strominger~\cite{andy}. More recently, the logarithmic
contributions to the black hole entropy have been discussed in
the
literature~\cite{carlip4,jing}. One would then have to assess the
relative size of our logarithmic term compared to those given in
these works.

\vspace{0.5in}
This work was supported, in part by the Department of Energy
under the contract number DOE-FG02-84ER40153. We would like to
thank Philip Argyres and Alex Lewis for reading the manuscript
and suggesting improvements.

\end{document}